\documentclass[a4paper,10pt]{article}

\usepackage{tensor}
\usepackage{amsmath}
\usepackage{amsfonts}
\usepackage{amssymb}
\usepackage{pstricks}
\usepackage{mathrsfs}

\title{Space--time fluctuations and the spreading of wavepackets}
\author{Ertan G\"okl\"u$^1$, Claus L\"ammerzahl$^1$, Abel Camacho$^2$ and Alfredo Macias$^2$ \\ 
$^1$ ZARM, University Bremen, Am Fallturm, 28359 Bremen, Germany \\ 
$^2$ Departamento de Fisica, Universidad Autonoma Metropolitana--Iztapalapa, \\
A.P. 55-534, Mexico D.F. 09340, Mexico}

\begin{document}

\maketitle

\begin{abstract}
Using a density matrix description in space we study the evolution of wavepackets in a fluctuating space--time background. We assume that space--time fluctuations manifest as classical fluctuations of the metric. From the non--relativistic limit of a non-minimally coupled Klein--Gordon equation we derive a Schr\"odinger equation with an additive gaussian random potential. This is transformed into an effective master equation for the density matrix. The solutions of this master equation allow to study the dynamics of wavepackets in a fluctuating space--time, depending on the fluctuation scenario. We show how different scenarios alter the diffusion properties of wavepackets. 
\end{abstract}

\section{Introduction}

The search for a quantum theory of gravity is still a work in progress and leaves many open questions. Currently there is no final version of a quantum gravity theory but it is expected that one of its consequences is the appearance of some kind of space--time foam. This can be thought of as space--time fluctuations which could manifest classically in the low--energy limit as fluctuations of the metric or of the connection.

In the most simple version one can think of a Minkowskian background on which small space--time dependent metrical fluctuations are imposed. In this context it has been shown for quantum mechanical systems that such a scenario has the following consequences: (i) we obtain a modified inertial mass and, thus, an apparent violation of the weak equivalence principle \cite{JaekelReynaud94,Camacho2003GRG,GoeklueLaemmerzahl08}. (ii) quantum systems will suffer decoherence induced by such quantum gravity induced space--time fluctuations which were discussed in \cite{PowerPercival00}-\cite{BreuerGoeklueLaemml09}. In the context of the propagation of light (iii) fluctuating light cones and (iv) angular blurring were discussed in \cite{Ford05}. Since space--time fluctuations can also lead effectively to non--localities they can manifest themselves as (v) modified dispersion relations \cite{DeformedDispers0}-\cite{DeformedDispers3}.

(vi) higher-order time derivatives  can appear in the context of equations of motions of higher order \cite{LamRade09} leading to non--localities w.r.t. the time variable. It might be expected that these terms appear as a consequence of quantum gravity scenarios. 

In this paper we investigate a further consequence of the model worked out in \cite{GoeklueLaemmerzahl08}. We consider the effect of space--time fluctuations on the evolution of wavepackets and calculate their modified evolution in terms of the mean--squared displacement and higher--order moments. The modification depends on the fluctuation scenario characterized by a spatial correlation function. Such kind of evolution equations have been discussed in \cite{Heinrichs92}-\cite{Jayannavar93} in the context of quantum diffusion of particles in dynamical disordered continua. In this work we will apply these methods to our model of space--time fluctuations and show that also in our case one obtains modified wavepacket dynamics dependent on the properties of the stochastic model.

\section{Modified quantum dynamics}

In our model space--time is regarded in the low energy limit as a fluctuating entity which consists of a classical fixed background on which Planck scale fluctuations are imposed. We assume that they appear as classical fluctuations of space--time and model this as perturbations of the metric up to second order 
\begin{eqnarray}
g_{\mu\nu}(\mathbf{x},t)& = & \eta_{\mu\nu}+h_{\mu\nu}(\mathbf{x},t),  \\ \label{modmetric1}
g^{\mu\nu}(\mathbf{x},t)& = &\eta^{\mu\nu} - h^{\mu\nu}(\mathbf{x},t) + \tilde{h}^{\mu\nu}(\mathbf{x},t), \label{modmetric1a}
\end{eqnarray}
where $|h_{\mu\nu}| \ll 1$. In this paper greek indices run from 0 to 3 and latin indices from 1 to 3.
The second order perturbations are given by $\tilde{h}^{\mu\nu}=\eta_{\kappa\lambda}h^{\mu\kappa}h^{\lambda\nu}$ where indices are raised and lowered with $\eta_{\mu\nu}$ and $\eta^{\mu\nu}$.

We consider a scalar field $\phi(x)$ non--minimally coupled to gravity as given by the action
\begin{equation}
S = \frac{1}{2}\int d^4x \sqrt{g} \left(g^{\mu\nu} \partial_\mu\phi(x)^* \partial_\nu\phi(x) - \left(\frac{m^2c^2}{\hbar^2}+\xi R(x)\right) \phi(x)^* \phi(x)\right) \, ,
\end{equation}
where $g=-{\rm det} g_{\mu\nu}$ is the determinant of the metric and $x=(\mathbf{x},t)$. Here $R(x)$ is the Ricci scalar and $\xi$ a numerical factor. Variation of the action with respect to $\phi(x)$ yields the Klein--Gordon equation non--minimally coupled to the gravitational field
\begin{eqnarray}
g^{\mu\nu}D_{\mu}\partial_{\nu}\phi(x)-\left(\frac{m^2c^2}{\hbar^2}+\xi R(x)\right)\phi(x)=0,
\end{eqnarray}
where $D_{\mu}$ is the covariant derivative based on the metric $g_{\mu\nu}$.

There are three values of $\xi$ which are of particular interest \cite{BirDav1982}: (i) $\xi=0$ is the minimally coupled case, (ii) $\xi=\frac{1}{6}$ is required by  conformal coupling and can also be derived from the requirement that the equivalence principle is valid for the propagation of scalar waves in a curved space--time \cite{SonegoFaraoni93}. (iii) $\xi = \frac{1}{4}$ originates from the squaring the Dirac equation. 
 
\subsection{Non--relativistic limit}

In the next step we calculate the non--relativistic approximation of this equation by expanding the wavefunction $\phi(\mathbf{x},t) = e^{iS(\mathbf{x},t)/\hbar}$ in powers of $c^2$ 
\begin{equation}
S(\mathbf{x},t)=S_0(\mathbf{x},t) c^2 + S_1(\mathbf{x},t) + S_2(\mathbf{x},t) c^{-2} + \ldots , \label{expansion1}
\end{equation}
according to the scheme worked out by Kiefer and Singh \cite{KieferSinghPRD1991}.
To orders $c^4$ and $c^2$ in the series expansion of the Klein--Gordon equation the results are identical to the case where minimal coupling is assumed \cite{GoeklueLaemmerzahl08}.
For obtaining the Schr\"odinger equation without relativistic corrections in the case of non-minimally coupling we give a sketch of the calculations which are done here analogously to \cite{GoeklueLaemmerzahl08}. First one has to derive the equation of motion to order $c^0$, where non-hermitian terms appear in the Hamiltonian. In order to get rid of them one can choose a modified scalar product in curved space or alternatively keep the usual `flat' euclidean scalar product and transform the operators and the wavefunction accordingly \cite{LamPhysLetA1995}.
We proceed with the latter possibility and finally arrive at
\begin{eqnarray}
i \hbar \partial_t \psi&=& -(^{(3)}g)^{1/4}\frac{\hbar^2}{2m} \Delta_{{\rm cov}}\left((^{(3)}g)^{-1/4}\psi\right)-\frac{\hbar^2}{2m}\xi R(x)\psi + \frac{m}{2} g^{00}\psi\nonumber \\
& & +\frac{1}{2}\left\lbrace i \hbar \partial_i, g^{i0}\right\rbrace \psi\label{hamtrans}.
\end{eqnarray}
Here $\Delta_{{\rm cov}}$ represents the Laplace--Beltrami operator, $\lbrace \cdot , \cdot \rbrace$ is the anticommutator, $^{(3)}g$ is the determinant of the 3--metric $g_{ij}$ and $g^{00}$ includes the Newtonian potential $U(\mathbf{x})$. The Hamiltonian is manifest hermitian with respect to the chosen `flat' scalar product. 

\subsection{The effective Schr\"odinger equation}

In \cite{GoeklueLaemmerzahl08} it is assumed that the particle described by the modified Schr\"odinger equation has its own finite spatial resolution scale. Consequently, only an averaged influence of space--time fluctuations can be detected. This is quantified by means of the spatial average of the modified Hamiltonian over the particle scale leading to an effective Hamiltonian
\begin{equation}
H =-\frac{\hbar^2}{2m}\left(\left(\delta^{ij}+\alpha^{ij}(t)\right)\partial_i \partial_j\right) -m U(\mathbf{x})  \, ,
\label{effhamiltonian}
\end{equation}
where the tensorial function $\alpha^{ij}(t)$ consists of the spatial average of squares of the metrical fluctuations.
The tensorial function $\alpha^{ij}(t)$ can be split into a time--average part and a fluctuating part 
\begin{equation}\label{alphasplit}
\alpha^{ij}(t) = \tilde\alpha^{ij} + \gamma^{ij}(t) \quad
{\rm with} \quad \langle \gamma^{ij}(t) \rangle = 0 .
\end{equation}
It has been schown that the part $\tilde{\alpha}^{ij}$ leads to a renormalized inertial mass implying an apparent breakdown of the Weak Equivalence Principle  \cite{GoeklueLaemmerzahl08} whereas the fluctuating part leads to a decay of  coherences in the energy representation \cite{BreuerGoeklueLaemml09}. If we include non--minimal coupling then a spatially averaged Ricci scalar $\left<R\right>_{V}$ appears leading to an additional term in the perturbation hamiltonian
\begin{equation}
 H_p=-\frac{\hbar^2}{2m}\left(\gamma^{ij}\partial_i \partial_j + \xi \left<R\right>_{V}\right),
\end{equation}
which commutes with the density operator $\tilde{\rho}$ of the effective master equation in \cite{BreuerGoeklueLaemml09}. Consequently, this extra term does not change the decoherence properties of quantum systems.

In the following sections we will derive the modifications of the dynamics of wave packets as induced by the fluctuating space--time metric. This requires an appropriate approximation of the kinetic term. 

\subsection{Approximation of the kinetic term}

Now we consider the full hamiltonian (\ref{hamtrans}) and analyze the kinetic term which is composed of the Laplace--Beltrami operator and the fluctuating metric quantities. It leads to the expression
\begin{eqnarray}
(^{(3)}g^{1/4})\Delta_{{\rm cov}}(^{(3)}g^{-1/4})& = & g^{ij}\left((^{(3)}g^{1/4})\partial_i\partial_j (^{(3)}g^{-1/4})-\frac{1}{2}\partial_i \ln{\sqrt{(^{(3)}g)}}\partial_j \ln{\sqrt{(^{(3)}g)}}\right)\nonumber \\
& & -\frac{1}{2}\partial_i g^{ij}\partial_j \ln{\sqrt{(^{(3)}g)}}+ g^{ij}\partial_i\partial_j+\partial_i g^{ij}\partial_j.
\end{eqnarray}
All contributions of the kinetic part as well as those terms in the anticommutator of equation (\ref{hamtrans}) which do not include derivatives acting on the wavefunction give a modified stochastic scalar interaction term 
\begin{eqnarray}
V(\mathbf{x},t) & = & -\frac{\hbar^2}{2m}\left( \xi R+g^{ij}\left((^{(3)}g^{1/4})\partial_i\partial_j (^{(3)}g^{-1/4})\right)\right) \nonumber \\
 & & -\frac{\hbar^2}{2m}\left(-\frac{1}{2}\partial_i \ln{\sqrt{(^{(3)}g)}}\partial_j \ln{\sqrt{(^{(3)}g)}}-\frac{1}{2}\partial_i g^{ij}\partial_j \ln{\sqrt{(^{(3)}g)}}\right) \nonumber\\
 & & +\frac{i\hbar}{2}\partial_i g^{i0} \label{randpotentialfull}
\end{eqnarray}
appearing in the Schr\"odinger equation
\begin{eqnarray}
i \hbar \partial_t \psi&=& -\frac{\hbar^2}{2m} \left( g^{ij}(\mathbf{x},t)\partial_i\partial_j+\partial_i g^{ij}\partial_j\right) \psi + i\hbar g^{i0}\partial_i \psi + V(\mathbf{x},t)\psi \label{schroed1},
\end{eqnarray} 
where for simplicity the Newtonian potential $g^{00}$ is not included henceforth. Our conclusions will not be affected by taking a Newtonian gravitational field in to account. 

Now we will employ some approximations which allows to use established methods for the calculation of wavepacket dynamics. We assume that space--time fluctuations act on a space--time scale which is much smaller than typical length- and timescales on which the wavefunction $\psi(\mathbf{x},t)$ varies. Therefore the amplitude of the terms containing derivatives of the fluctuating quantities ($h^{ij}$ and the trace $h$) dominate in the Schr\"odinger equation (\ref{schroed1}) compared with the magnitude of the other fluctuating quantities. This allows us to state that the main contributions from spacetime noise stem from terms containing products of first derivatives of the fluctuation quantities . 
In short notation  
\begin{eqnarray}
|\partial h \partial h| > (h, \tilde{h}) \partial \partial, \quad |\partial h \partial h| > h \partial h \partial \quad \text{and} \quad |\partial h \partial h| > \partial \tilde{h} \partial,
\end{eqnarray}
where for simplicity we omitted indices.
Therefore we can neglect the terms $\left(\tilde{h}^{ij}-h^{ij}\right)\partial_i \partial_j$, $\partial_i g^{ij}\partial_j$ and $\partial_i g^{i0}$. 
In addition, we note that in the following derivation of the averaged master equation only second moments of $V$ will appear, see Eq. (\ref{corr3}). This would lead to third and fourth order terms if we incorporate second order fluctuation terms (third order terms vanish in the average because of the gaussian property which we will specify later). Therefore we can neglect second order terms already on the level of the Schr\"odinger equation leading to
\begin{equation}
 i \hbar \partial_t \psi=-\frac{\hbar^2}{2m}\Delta \psi +V(\mathbf{x},t) \psi \label{schroed2},
\end{equation}
where $\Delta$ is the Laplace operator in cartesian coordinates and where the random scalar interaction (\ref{randpotentialfull}) now reads
\begin{eqnarray}
V(\mathbf{x},t)& = & -\frac{\hbar^2}{2m}\left( \xi R+\delta^{ij}\left((^{(3)}g^{1/4})\partial_i\partial_j (^{(3)}g^{-1/4})\right)\right). 
\end{eqnarray}
To first order in the perturbation metric this reduces to
\begin{eqnarray}
V(\mathbf{x},t)& = & -\frac{\hbar^2}{2m}\left( \xi R-\frac{1}{4}\Delta h\right), \label{randpot1}
\end{eqnarray}
where $h$ is the first-order spatial part of the trace $\text{tr}(g_{\mu\nu})$.
To first order the Ricci scalar is given by
\begin{eqnarray}
 R(\mathbf{x},t) & = & \partial_l \partial_k h^{lk}(\mathbf{x},t)-\Delta h(\mathbf{x},t) \label{ricci1}\, ,
\end{eqnarray}
Note that only spatial components are present which is an effect of the non-relativistic approximation (\ref{expansion1}).
Compared to the second term in the fluctuating potential (\ref{randpot1}) the non-minimal coupling term $\xi R$ has the same structure and the same order of magnitude. Thus this contribution to the Schr\"odinger Hamiltonian and to the analysis of space--time fluctuations cannot be neglected in the context of our model and must be taken into consideration in the following calculations.

\subsection{Stochastic properties}

Now we have to specify the statistical properties of the fluctuating potential $V(\mathbf{x},t)$ as we like to derive an effective equation. 
Owing to the fluctuating metric this term will be interpreted as a gaussian random function $V(\mathbf{x},t)$ dependent on the quantities $h_{\mu\nu}(\mathbf{x},t)$. 
We will now continue the calculations for the one dimensional case as in the next section the effective master equation is derived for this specific case and from now on the spatial coordinate is denoted by $x$.
This yields for the fluctuating potential
\begin{eqnarray}
 V(x,t) & = &-\frac{\hbar^2}{2m}\left(\xi \partial_x^2 h_{11}(x,t)-\partial_x^2 h_{11}(x,t)\left(\xi+\frac{1}{4}\right)\right) \\
& = & \frac{\hbar^2}{8m} \partial_x^2 h_{11}(x,t).
\end{eqnarray}
As the quantities of interest are the metric perturbation terms, we choose the following statistical conditions
\begin{eqnarray}
\left<h_{11}(x,t)\right> & = &0 \label{corr1}\\
C(x,x',t,t')\equiv\left<h_{11}(x,t)h_{11}(x',t')\right> & = & h_0^2 \delta(t-t')C( x-x') \, ,\label{corr2}
\end{eqnarray}
where $h_0$ is the strength of the fluctuations of $h(x,t)$. It has been shown by Heinrichs \cite{Heinrichs92, {Heinrichs96}} that the dynamics of wavepackets are unaffected by temporal correlations of the random function $h(x,t)$ if they are sufficiently small. This also applies to our model of space--time fluctuations because it is commonly believed that quantum gravitational induced fluctuations appear on typical scales given by the Planck time $\tau_p$ and Planck length $l_p$. Therefore we have chosen a $\delta$--correlation with respect to time. However, at the moment the spatial correlation function $C(x-x')$ will be left unspecified.
For the following calculations it is convenient to introduce the correlation function 
 \begin{eqnarray}
 \left<V(\mathbf{x},t)V(\mathbf{x}',t')\right> & = & V_0^2 \delta(t-t') g(x-x') \label{corr3},
 \end{eqnarray}
where $g(x-x')=\partial_x^2 \partial_{x'}^2 C(x-x')$ and we defined $V_0=\frac{\hbar^2}{8 m}h_0$.

As this is a model for space--time fluctuations and because of the lack of a complete understanding of the microscopic structure of space--time one has the freedom to characterize the statistical properties of the terms $h$ with a variety of stochastic models given here by the correlator $C(x-x')$.

\section{Dynamics of the density matrix}

\subsection{Effective master equation}

Most physical quantities of interest -- in our case the mean square displacement -- can be conveniently calculated in terms of the reduced density matrix $\left<\rho(x',x,t)\right>$ and its evolution equation. For the sake of simplicity we restrict all the calculations to the one--dimensional case. In the following we pursue closely the strategy and calculation techniques realized in \cite{Jayannavar93, {Jayannavar82}}, which lead to an exact solution of the master equation and thus for the mean-squared displacement of a wavepacket. In addition to this we clarify some derivations and assumptions made in \cite{Jayannavar93} and apply the results to our model of space--time fluctuations.

We define the density operator 
\begin{equation}
 \rho(x',x,t)=\psi^*(x',t)\psi(x,t)
\end{equation}
and set up the master equation according to
\begin{eqnarray}
\partial_t \rho(x',x,t)& = &\partial_t (\psi^*(x',t)) \psi(x,t) + \psi^*(x',t) \partial_t \psi(x,t)\\
& = & \frac{i \hbar}{2m}\left(\frac{\partial^2}{\partial x^2}-\frac{\partial^2}{\partial x'^2}\right)\rho(x',x,t)+\frac{i}{\hbar}\left(V(x',t)-V(x,t)\right)\rho(x',x,t). \label{mastereq}
\end{eqnarray}
The formal solution of this master equation reads
\begin{eqnarray}
\rho(x',x,t)& = &\rho(x',x,t=0)+\frac{i \hbar}{2m}\int^t_0 dt' \left(\frac{\partial^2}{\partial x^2}-\frac{\partial^2}{\partial x'^2}\right)\rho(x',x,t') \nonumber \\
& & + \frac{i}{\hbar}\int^t_0 dt'\left(V(x',t')-V(x,t')\right)\rho(x',x,t').\label{formsolmastereq}
\end{eqnarray}
In the next step we calculate the effective equation of motion by applying the average over the fluctuations. Special care must be taken for the terms which are products of the fluctuating potential $V(x,t)$ and the density operator. This is due to the fact that the average of the product $\left<V(x,t)\rho(x',x,t)\right>$ generally does not factorize because the density operator $\rho(x',x,t)$ is a functional of the fluctuation potential $ V(x,t)$, hence $\rho[V]$. 

We can handle this by using the fact that we have chosen gaussian fluctuations (see the statements preceding Eqns.~(\ref{corr1}) and (\ref{corr2})). A gaussian process is characterized in functional form by the Novikov theorem \cite{Novikov64} valid for processes obeying $\left<V(x,t)\right>=0$ (here:  $\partial_x^2 \left< h_{11}(x,t)\right>=0$), which yields in our case
\begin{eqnarray}
 \left<V(x,t)\rho(x',x,t)\right> & = & \int dt'' \int dx'' \delta(t-t'') g(x-x'')\left<\frac{\delta  \rho(x',x,t)}{\delta V(x'',t'')}\right> \label{novikov1},\\
\left<V(x',t)\rho(x',x,t)\right> & = & \int dt'' \int dx''\delta(t-t'') g(x'-x'')\left<\frac{\delta  \rho(x',x,t)}{\delta V (x'',t'')}\right>\label{novikov2},
\end{eqnarray}
where $g$ is the correlation function introduced in equation (\ref{corr3}). 

The Novikov theorem follows from a functional Taylor series expansion of $\rho[V]$ and utilizing the gaussian property of $h_{11}(x,t)$ which leads to a factorization of moments \cite{Sancho82}.
The functional derivative can be calculated using the formal solution of the master equation (\ref{formsolmastereq}) and yields
\begin{eqnarray}
 \left<\frac{\delta  \rho(x',x,t)}{\delta V(x'',t'')}\right> & = &\frac{i}{\hbar} \int_0^t dt' \rho(x',x,t')\left(\delta(x-x'')\delta(t'-t'')-\delta(x'-x'')\delta(t'-t'')\right) \nonumber \\
& = & \frac{i}{2\hbar} \rho(x',x,t'')\left(\delta(x-x'')-\delta(x'-x'')\right).
\end{eqnarray}
This result used in (\ref{novikov1}) yields
\begin{eqnarray}
\left<V(x,t)\rho(x',x,t)\right> & = & \frac{i V_0^2}{2\hbar}\int dx''  g(x-x'')\left<\rho(x',x,t)\right> \left(\delta(x-x'')-\delta(x'-x'')\right)\nonumber\\
& = & \frac{i V_0^2}{2\hbar}\left<\rho(x',x,t)\right>\left(g(0)-g(x-x')\right),
\end{eqnarray}
and analogously, from equation (\ref{novikov2}),
\begin{eqnarray}
\left<V(x',t)\rho(x',x,t)\right> & = & \frac{i V_0^2}{2\hbar}\left<\rho(x',x,t)\right>\left(g(x-x')-g(0)\right).
\end{eqnarray}

Combining both results and inserting this into the averaged Equation (\ref{mastereq}) we obtain the effective master equation for the reduced density matrix $\left<\rho(x',x,t)\right>$
\begin{eqnarray}
\frac{\partial }{\partial t}\left<\rho(x',x,t)\right>  & = & -\frac{i \hbar}{2m}\left(\frac{\partial^2}{\partial x'^2}-\frac{\partial^2}{\partial x^2}\right)\left<\rho(x',x,t)\right> \nonumber \\
& & - \frac{V_0^2}{\hbar^2}\left(g(0)-g(x-x')\right)\left<\rho(x',x,t)\right>.
\end{eqnarray}

\subsection{Solution of the effective master equation}

In order to get rid of the time dependence we perform a Laplace transformation according to
\begin{equation}
 r(x',x,s)=\mathcal{L}[\left<\rho(x',x,t)\right>]=\int_0^{\infty} dt \left<\rho(x',x,t)\right> e^{-st},
\end{equation}
where the Laplace variable $s$ satisfies ${\rm Re}(s)>0$.
The time derivative becomes
\begin{equation}
\mathcal{L}[\partial_t \left<\rho(x',x,t)\right>]=s \mathcal{L}[\left<\rho(x',x,t)\right>]-\left<\rho(x',x,t=0)\right> 
\end{equation}
and the Laplace transformed master equation reads
\begin{eqnarray}
\frac{i \hbar}{2m}\left(\frac{\partial^2}{\partial x'^2}-\frac{\partial^2}{\partial x^2}\right)r(x',x,s) + \left(s+\frac{V_0^2}{\hbar^2}\left(g(0)-g(x-x')\right)\right)r(x',x,s)=\left<\rho(x',x,t=0)\right>,
\end{eqnarray}
where $\left<\rho(x',x,t=0)\right>$ is subject to initial conditions.
For practical reasons we introduce new coordinates according to $X=x+x'$ and $Y=x-x'$ which yields
\begin{eqnarray}
 \frac{2 i \hbar}{m}\frac{\partial^2}{\partial X \partial Y} r(X,Y,s)+\left(s+\frac{V_0^2}{\hbar^2}\left(g(0)-g(Y)\right)\right)r(X,Y,s)=\left<\rho(X,Y,t=0)\right>.\label{mastereq2}
\end{eqnarray}
This is the partial differential equation (PDE) which has to be solved. 

By using the averaging techniques to derive the effective equation (\ref{mastereq2}) we converted the stochastic PDE (\ref{mastereq}) to an 'ordinary' PDE. Note that the information about the fluctuation scenario is encoded in the correlation function $g(Y)$ and the amplitude $V_0$. 

We rewrite the master equation (\ref{mastereq2}) as an ordinary, inhomogeneous first-order differential equation by applying a Fourier transformation w.r.t. the variable $X$
\begin{equation}
 \left(\mathcal{F}_{X}r\right)(K,Y,s)=R(K,Y,s)=\int^{\infty}_{-\infty} dX e^{-iKX} r(X,Y,s) \, .
\end{equation}
This gives
\begin{equation}
 \partial_Y R(K,Y,s) - \frac{m}{2\hbar K} \left(s+\frac{V_0^2}{\hbar^2}\left(g(0)-g(Y)\right)\right)R(K,Y,s)=- \frac{m}{\hbar K}R(K,Y,t=0) \, ,
\end{equation}
where $R(K,Y,t=0) =\left(\mathcal{F}_{X}\left<\rho\right>\right)(K,Y,t=0)$ is the Fourier transform of the inhomogeneous part of the ordinary differential equation.

This equation can easily be integrated and yields
\begin{eqnarray}
R(K,Y,s)& = & \exp{\left(\frac{m s Y}{2\hbar K}+G(Y) \right)}R(K,Y_0,s)-\exp{\left(\frac{m s Y}{2\hbar K}+G(Y) \right)}\times \label{solution1} \\
& & \times\frac{m}{\hbar K}\int_0^Y dY' \exp{\left(-\frac{msY'}{2\hbar K}\right)} R(K,Y',t=0) \exp{\left(-\frac{mG(Y')}{2 \hbar K}\right)}, \nonumber 
\end{eqnarray}
where the term $R(K,Y_0,s)=R(K,Y=0,s)$ represents the initial value of the solution $R(K,Y,s)$ and we defined $G(Y)=\frac{V_0^2}{\hbar^2}\int^Y_0 dY' \left(g(0)-g(Y')\right)$.

\subsection{Mean squared displacement}

The mean squared displacement of the particle is given by 
\begin{equation}
\sigma_x^2(t) = \int^{\infty}_{-\infty} dx\, x^2 \, \lim_{x\rightarrow x'}\left<\rho(x',x,t)\right> 
\end{equation}
and can be expressed in Fourier space as 
\begin{eqnarray}
\sigma_x^2(t) =-\frac{1}{8}\frac{\partial^2}{\partial K^2}R(K,Y_0,t)\Big|_{K=0}.\label{msdisplacement}
\end{eqnarray}
Therefore the quantity of interest which has to be calculated is the initial value of $R$ at $Y=0$, corresponding to $x=x'$ and hence to the diagonal element of the density matrix $\rho(x',x,t)$. 

The solution $R(K,Y,s)$ is a  quadratically integrable function and must vanish for $Y\rightarrow \infty$. Therefore the r.h.s. of equation (\ref{solution1}) must be zero in this limit leading to
\begin{eqnarray}
R(K,Y_0,s) =  \frac{m}{\hbar K}\int_0^{\infty} dY' \exp{\left(-\frac{msY'}{2\hbar K}\right)} R(K,Y',t=0) \exp{\left(-\frac{mG(Y')}{2 \hbar K}\right)}.
\end{eqnarray}
We make the substitution $\tau=\frac{m}{2\hbar K}Y$ (the dimension of this quantity is in fact that of time) and note that the r.h.s. represents a Laplace transformation  
\begin{equation}
 R(K,Y_0,s)=2 \int_0^{\infty} d\tau  \exp{\left(-s\tau\right)} R(K,\frac{2 \hbar K}{m}\tau,t=0) \exp{\left(-G(\tau)\right) },
\end{equation}
if $\tau$ is positive giving $K \rightarrow |K|$.
This equation can be inverted leading to the solution
\begin{equation}
R(K,Y_0,\tau)= 2 \, R(K,\frac{2 \hbar |K|}{m}\tau,t=0)\exp{\left(-\frac{V_0^2}{\hbar^2}\int^{\tau}_{0} d\tau'  \left[g(0)-g\left(\frac{2 \hbar |K|}{m}\tau'\right)\right]\right)}.
\end{equation}
\\
Inserting this expression into equation (\ref{msdisplacement}) we get for the mean squared displacement
\begin{equation}
 \sigma_x^2(t) = \frac{1}{4}\,{{\rm e}^{-G (Y)}}R_0(Y) \left(
\partial_k ^{2}G(Y) - \left(\partial_k G(Y)  \right) ^{2} +2\, \partial_k G(Y) \partial_k \ln{R_0(Y)}-\partial^2_k \ln{R_0(Y)}-(\partial_k \ln{R_0(Y)})^2  \right)\Big|_{K=0}, \label{msdispplacementgeneral}
\end{equation}
where $Y=2 \frac{\hbar |K|}{m}\tau$ and $R_0(Y)=R(K,\frac{2 \hbar |K|}{m}\tau,t=0)$.
\\

Note that the unperturbed solution is given by the last two logarithmic terms, where for $K=0$ we get $\exp{(-G (Y))}R_0(Y)=1$, if the initial state $R_0$ is given by a gaussian wavepacket. The information about the fluctuation scenario is given by the correlation function appearing in $G(Y)$. In our case we regard even correlation functions $G(Y)=G(-Y)$ excluding a preferred direction in space. For our purposes it is assumed to have the form $G(Y^{2n})$, for $n \in \mathbb{N}$ avoiding the introduction of the modulus of the argument $Y$ as the correlation function must be differentiable and analytic at the point $Y=0$. In this case the first derivative $\partial_k G(Y)$ vanishes in the limit $K=0$, therefore modifications to the free (unperturbed) mean squared displacement come from the second derivative $\partial^2_k G(Y)$.

\subsection{Gaussian correlator}

With this result (\ref{msdispplacementgeneral}) we can now calculate the mean squared displacement of a wavepacket.
In the preceding discussion we imposed several restrictions and assumptions for the argument of the correlation function $C(Y)$. For this reason and as a first starting point we choose a gaussian correlator
\begin{equation}
 C(Y)=\frac{1}{\sqrt{2\pi}a}\exp{\left(-\frac{Y^2}{a^2}\right)} \label{corrspat1},
\end{equation}
where $a$ represents a finite correlation length and we choose for the initial conditions a gaussian wavepacket
\begin{eqnarray}
R(K,Y,t=0) & = & \mathcal{F}_{X}\left[ \frac{1}{(2 \pi)^{1/2}\sigma }\exp{\left(-\frac{X^2+Y^2}{8\sigma^2}\right)}\right]\\
& = & 2\, \exp{\left(-\left(\frac{Y^2}{8 \sigma^2}+2K^2\sigma^2\right)\right)}.
\end{eqnarray}
We set $Y=2 \frac{\hbar |K|}{m}\tau$ and the mean squared displacement yields
\begin{eqnarray}
 \sigma_x^2(t) = \sigma^2_x(0)+ \frac{\hbar^2}{4m^2\sigma^2_x(0)}t^2 + \frac{5 V_0^2}{\sqrt{2\pi}m^2 a^7} t^3. \label{msd1}
\end{eqnarray}
The first two terms correspond to the expression that is obtained from the free Schr\"odinger equation whereas the last term accounts for superdiffusive \footnote{The nomenclature is at follows: For the mean-squared displacement $\left<x^2\right>\propto t^{\nu}$ the exponent $\nu=1$ renders diffusive motion, $\nu=2$ is a ballistic motion and $\nu=3$ denotes superdiffusion.} behavior. The asymptotic behavior for $t\rightarrow \infty$ is dominated by the $t^3$ dependence 
\begin{equation}
 \sigma^2_x(t) = \sigma^2_x(0) +\frac{5 V_0^2}{\sqrt{2\pi}m^2 a^7} t^3 , \qquad \text{for } t\rightarrow \infty.\label{msd3}
\end{equation}
We can state that for long evolution times the wavepacket undergoes superdiffusion if the spatial correlation properties of space--time fluctuations are given by (\ref{corrspat1}). The effect of this special model of spacetime fluctuations is such that the evolution of quantum particles is superdiffusive.

The inclusion of a white noise scenario by setting $C(x-x')=\delta(x-x')$ is not feasible in this framework. The reason is that our equations are only valid for analytical correlation functions (analytic at $K=0$) but singular quantities  like $\delta(0)$ would appear in the expressions which were derived here. These expressions would render infinite momentum contributions which must be limited by a momentum--cutoff. In our framework this is motivated by a fundamental minimal length-scale which is given by the Planck length, since we consider possible effects emerging from quantum gravity. This leads to the desired momentum-cutoff and hence the Dirac $\delta$-function is not the appropriate choice in this context.   

However, a modified behavior for wavepackets was already sketched in \cite{BreuerGoeklueLaemml09} where a white noise scenario for the fluctuating phase of the wavefunctions was considered yielding
\begin{equation}
 \sigma^2_x(t) = \sigma^2_x(0)+\frac{\sigma_{px}(0)}{m}t+\frac{\sigma_p^2}{m^2}t^2+\frac{\sigma_p^2}{m^2}\tau_p t.
\end{equation}
The quantities appearing here are given by $ \sigma_{px}(t) = \langle p x + x p \rangle_t - 2\langle p \rangle_t \langle x \rangle_t$ and $\sigma^2_p(t)=\langle p^2 \rangle_t
- \langle p \rangle^2_t$ which is constant in time. The constant $\tau_p$ appearing in the dissipative term is the Planck time and given by the amplitude of the white noise scenario whereas the average is defined by the trace over the effective density operator $\langle A\rangle_t=\text{tr}\lbrace A \left<\rho(t)\right>\rbrace$.
However, for large times $t\gg \tau_p$ the quadratically increasing ballistic term dominates and the dissipative linear term can be neglected thus reproducing the behavior of a wavepacket described by a free Schr\"odinger equation. 

Therefore the result (\ref{msd3}) derived here by introducing a finite correlation length (gaussian correlator) represents a new effect emerging from our model for spacetime fluctuations. 

\subsection{Other correlators and higher order moments}

In general it is not possible to distinguish different fluctuation scenarios on the level of wavepacket dynamics only by the second moment (\ref{msdisplacement}). 
This does not suffice to fully characterize the modified wavepacket evolution what we demonstrate here by selecting the correlation function
\begin{equation}
  C(Y)= C_4\, \exp{\left(-\frac{Y^4}{4 a^4}\right)}, \label{corrspat2}
\end{equation}
where $C_4$ is a normalization factor. We arrive at
 \begin{eqnarray}
\sigma_x^2(t) = \sigma^{2}_x(0)+\frac{\hbar^{2}{t}^{2} }  {4{m}^{2}{\sigma^2_x(0)}}, \label{msd2}
 \end{eqnarray}
showing no superdiffusion and no deviation from the usual wavepacket dynamics.

Considering the case  $C(Y)=C_6\exp{\left(-\frac{Y^6}{a^6}\right)}$ leads us to
\begin{eqnarray}
 \sigma_x^2(t) = \sigma^2_x(0)+ \frac{\hbar^2}{4m^2\sigma^2_x(0)}t^2 + \frac{45 V_0^2 \Gamma(\frac{5}{6})}{\sqrt{2}\pi m^2 a^7} t^3 \label{msd4},
\end{eqnarray}
where $\Gamma$ is the Gamma function.
This expression has the same structure and asymptotic behavior like equation (\ref{msd1}). 

Generalizing these results to correlation functions $C(Y)=C_n \, \exp{ \left(-\left(\frac{Y}{\sqrt{2} a}\right)^{2n} \right)}$ \cite{Jayannavar93}, one yields again the expression (\ref{msd2}) for all $n \in \mathbb{N}_{>1}\backslash \left\lbrace 3 \right \rbrace $.
\\
In order to deal with this fact one must consider higher-order moments $\left<x^{2n}(t)\right>$ with $n \in \mathbb{N}$ and $n > 1$. 

For correlation functions having exponents of order $2n$ one must include moments up to the order $2n$ to fully characterize the response of the wavepacket to space--time fluctuations, where the higher order moments are given by
\begin{equation}
 \left<x^{2n}(t)\right>\propto \frac{\partial^{2n}}{\partial K^{2n}}R(K,Y_0,\tau)|_{K=0}.
\end{equation}
For example we get for the correlator (\ref{corrspat2}) the asymptotic ($t\rightarrow \infty$) fourth moment 
\begin{equation}
\left<x^4(t)\right> \approx 96 \sigma^4_x(0) + \frac{V_0^2\hbar^2}{  \pi m^4 a^9}t^5.
\end{equation}

Thus, in order to be able to distinguish between different stochastic scenarios for spacetime fluctuations one must characterize the wavepacket spreading by the inclusion of higher order moments.

\subsection{Connection to holographic noise}

As a side remark we want to argue whether and in which way holographic noise would affect the wavepacket propagation. While simple estimates lead to a linear scaling of the measurement uncertainty with the Planck length $l_p$, effects from space--time fluctuations in a holographic fluctuation scenario are {\it amplified}. 

The holographic principle states that the number of degrees of freedom of a region of space--time is bounded by the area of the region in Planck units $l_p^2$. Alternatively we can state that although the world appears to have three spatial dimensions, its contents can be encoded on a two-dimensional surface, like a hologram. According to this principle length measurements of a distance $l$ exhibit an intrinsic uncertainty given by $\delta l \geq (l\, l_p^2)^{1/3}$, where the exponent $2/3$ for the Planck length $l_p$ is characteristic for holographic noise  \footnote{Here the Planck length is assumed to be the fundamental minimal length.}. 
It can be shown \cite{JackNg2002}, \cite{JackNg2003} that this translates to fluctuations of the metric 
\begin{equation}
 \delta g_{\mu\nu} \geq (l_p / l)^{2/3} a_{\mu\nu},
\end{equation}
where $a_{\mu\nu}$ is a tensor of order $\mathcal{O}(1)$.
This corresponds to a power spectral density (PSD) of 
\begin{equation}
S(f)=f^{-5/6}(c l^2_p)^{1/3},
\end{equation}
where $f$ is the frequency and $c$ is the speed of light. We write this in terms of a PSD in momentum space obtaining
\begin{equation}
S(k)=k^{-5/6} l_p^{2/3},
\end{equation}
where $k$ is the wavenumber. It can be readily checked that this expression leads to $\delta l = (l\, l_p^2)^{1/3}$ by calculating $(\delta l)^2=\int^{k_{max.}}_{1/l} dk \left(S(k)\right)^2$, where $k_{max.}$ is subject to a cut-off given by the Planck length $l_p$ and the length $l$ is given by the experiment.
This shows that -- by means of a correlation function -- a finite, nonvanishing correlation length is involved where the correlator scales with $ l_p^{2/3}$ modifying the wavepacket evolution accordingly. The $l_p^{2/3}$-dependence distinguishes the holographic noise scenario from the other fluctuations scenarios discussed in the previous sections.

\subsection{Experimental suggestions}
For the experimental observation of the predicted effects on wavepacket dynamics it is necessary that long evolution times of quantum particles can be achieved. 
So far, it seems that no dedicated experiment concerning the spreading of quantum mechanical wavepackets has been performed. We suggest to test this effect.
In this context it is feasible that the gravitational field of earth is suppressed as best as possible to facilitate the free evolution of wavepackets. This goal can be reached by performing experiments in microgravity environments which is realized for example by the QUANTUS \cite{QUANTUS06} and PRIMUS projects.
The first already demonstrated that a BEC (Bose--Einstein-Condensate) in free fall can be realized with propagation times of about one second. In the PRIMUS project BECs will be used to realize a matter wave interferometer under microgravity conditions. We suggest that in both projects cold atom gases could in principle be used to study the long time behavior of wavepackets with high sensitivity. 

\section{Summary and conclusion}

We have calculated the mean squared displacement for a quantum particle subject to space--time fluctuations. 
As we have seen the properties of wavepacket dynamics are strongly dependent on the spatial correlation function. For two special cases we encounter superdiffusion of the wavepacket for the second moment on large timescales (characterized by the exponent of the time variable). For other correlators the second moment is unaffected and the unperturbed expression is recovered.  Therefore higher-order moments must be included in the analysis showing modifications of the unperturbed case. This leads also to a superdiffusive behavior and - in addition - allows to distinguish different stochastic models experimentally. 
For a white noise scenario we encounter a diffusive behavior for short timescales. On large timescales, compared to the typical evolution time of a wavepacket, the unperturbed case is recovered as the ballistic term dominates.  
Furthermore we discussed the influence of holographic noise on the wavepacket evolution and argued that it should be dependent on the Planck length appearing with the exponent $2/3$. 

Therefore, interesting effects representing deviations from the usual wavepacket dynamics can only be obtained by inclusion of general correlation functions having finite correlation lengths. In our approach the modifications on wavepacket dynamics depend only on the statistical fluctuation properties and the magnitude, whereas the latter is determined by the fluctuation strength independently of the statistical model.

Furthermore, we observe that in our model the strength of the modifications to the free Schr\"odinger equation for the minimally and the non--minimally coupled scenario have the same order of magnitude. However, the results for the one-dimensional case are independent of the non-minimal coupling term. For the 3-dimensional case we expect modifications coming from the non-minimally coupling terms but they will not affect our conclusions significantly. Since the structure of the equations remains the same and the form of the correlation functions are identical to those in one dimension we should not encounter major differences for the expressions for the wavepackat dynamics. The magnitude and the temporal behavior of the modifications should remain the same. Nevertheless, possible effects from anisotropic space--time fluctuations cannot be described in the one-dimensional case, of course, and therefore our conclusions do not include this possibility. 

We conclude that a possible experimental verification of our model of spacetime foam is feasible by means of experiments with cold atoms with long propagation times -- preferably in a microgravity environment.  This gives the opportunity of setting bounds for different scenarios of space--time fluctuations. 

\section*{Acknowledgements}

We like to thank \v{Z}. Marojevi\'{c} for fruitful discussions. All authors thank the German--Mexican DFG--CONACyT cooperative grants. EG gratefully acknowledges the support by the German Research Foundation (DFG) and the Centre for Quantum Engineering and Space--Time Research (QUEST), and CL the support by the German Aerospace Center (DLR) grant no. 50WM0534.

\end{document}